\documentclass[aps,pra,showkeys,superscriptaddress,notitlepage,11pt]{revtex4-1}
\usepackage{graphicx}  
\usepackage{float}
\usepackage[labelformat=simple]{subcaption} 
\usepackage{amsmath,amsfonts,amssymb,bbm,MnSymbol}   
\usepackage{ulem} 
\usepackage[breaklinks=true,colorlinks=true,linkcolor=blue,urlcolor=blue,citecolor=blue]{hyperref}
\usepackage{comment}
\usepackage{upgreek}
\usepackage{color}
\usepackage[makeroom]{cancel}
\pdfoutput=1
\usepackage{footnote}

\begin{document}

\title{Quantum and classical effects in a light-clock falling in Schwarzschild geometry}

\author{Maximilian P. E. Lock}
\affiliation{Department of Physics, Imperial College, London SW7 2AZ, United Kingdom}
\affiliation{Institute for Quantum Optics and Quantum Information (IQOQI), Austrian Academy of Sciences, A-1090 Vienna, Austria}
\author{Ivette Fuentes}\thanks{Previously known as Fuentes-Guridi and Fuentes-Schuller.}
\affiliation{School of Mathematical Sciences, University of Nottingham, University Park, Nottingham NG7 2RD, United Kingdom}

\begin{abstract}
Quantum theory and relativity offer different conceptions of time. To explore the conflict between them, we study a quantum version of the light-clock commonly used to illustrate relativistic time dilation. This semiclassical model combines elements of both theories. We show for Gaussian states of the light field that the clock time is independent of the initial state. We calculate the discrepancy between two such clocks when one is held in a gravitational field and the other is left to fall a certain distance. Contrasting our results with the case of pointlike observers in general relativity, as well as classical light-clocks, we find both quantitative and qualitative differences. We find that the quantum contribution to the discrepancy between the two clocks increases with the gravitational field strength, and results in a minimum resolution of the dropped clock (distinct from the quantum uncertainty in its measurement).
\end{abstract}

\maketitle

\vspace{-8mm}

\section{Introduction} \label{sIntro}

The original formulation of special relativity~\cite{einstein1905elektrodynamik} begins with a definition of simultaneity via the synchronisation of clocks by means of light signals sent between them. This definition includes a number of assumptions, namely that clocks are pointlike objects following well-defined trajectories in space, and can tick with arbitrary accuracy. To generalize to accelerated reference frames, one invokes the `clock hypothesis', according to which the apparent tick rate of a clock undergoing non-inertial motion equals that of an instantaneously comoving inertial one~\cite{rindler2006relativity}. A clock with this property is said to be `ideal'. It can be justified by noting that an observer can detect their own non-inertial motion via apparent forces, and hence a clock which deviates from ideality in a predictable manner can be combined with an accelerometer to give an ideal clock~\cite{sexl2001relativity}. Importing this into general relativity via the equivalence principle, one finds that an ideal clock measures a time proportional to the geometric length along its worldline~$\mathcal{C}$:
\begin{equation} \label{eGRPropTime}
\tau = \int_\mathcal{C} \sqrt{- g_{\mu \nu} \mathrm{d}x^\mu \mathrm{d}x^\nu},
\end{equation}
where $g_{\mu\nu}$ is the spacetime metric and we have chosen $c=1$.  

The construction of the theory of relativity by reference to the problem of synchronising clocks can be seen as an example of operationalism. According to this principle, concepts are to be defined with respect to the operations by which they must be measured~\cite{bridgman1927logic}. This principle also underlies the famous `light-clock' derivation of time dilation due to uniform relative motion, commonly attributed to Einstein, where one considers clocks constituted by two mirrors and a light ray propagating back and forth between them. For non-inertial motion, the orientation of the light-clock with respect to the direction of acceleration becomes relevant. In the case where the acceleration is parallel to the plane of the mirrors, the clock can be adapted by the addition of a third mirror~\cite{west2007light} in order to recover ideal behaviour. Light-clocks whose mirrors are perpendicular to the acceleration deviate from ideality~\cite{rindler2006relativity}, but in a manner that can be made arbitrarily small~\cite{fletcher2013light}. In any case, as noted above, such deviations can be measured concurrently and subtracted according to classical physics.

Maintaining this operationalist perspective while incorporating quantum theory, we are forced to reconsider the above reasoning. The clock cannot be assigned a definite trajectory in space, and the time elapsed according to the clock cannot be determined to arbitrary precision~\cite{braunstein1996generalized,peres1980measurement}. Furthermore, we must re-examine the possibility of constructing clocks satisfying the clock hypothesis, particularly in light of the Unruh~\cite{unruh1976notes} and dynamical Casimir effects~\cite{moore1970quantum,fulling1976radiation}, whereby non-inertial motion affects quantum states.

The study of quantum clocks has already proven fruitful in the non-relativistic setting, for example revealing problems arising from coupling the clock to a system whose evolution is to be tracked~\cite{peres1980measurement} (with significant consequences in the so-called Page and Wootters formalism~\cite{smith2017quantizing}), the role of energy and the size of the state space in limiting a clock's accuracy~\cite{buzek1999optimal}, and the extent to which quantum control can be autonomously implemented~\cite{woods2019autonomous}. Using a clock model where timekeeping is defined by the ability to consistently order events~\cite{rankovic2015quantum}, introducing the requirement that a clock runs autonomously has revealed the necessity of entropy generation in the process of keeping time~\cite{erker2017autonomous}. In a relativistic quantum setting, the role of clock-mass in limiting time measurements has been revealed~\cite{salecker1958quantum}, and deviations from clock ideality have been predicted in experiments using decaying particles to keep time~\cite{eisele1987on,lorek2015ideal,pierini2018can} as well as in a superconducting quantum interference device (due in part to relativistic particle creation)~\cite{lindkvist2014twin}. Moreover, an operational approach has also been used to shed light on the energy-time uncertainty relation in the context of the Bohr-Einstein gedankenexperiment~\cite{aharonov2000weighing}. The interplay of quantum interference and relativistic time dilation has also been studied~\cite{zych2011quantum,khandelwal2019general}. Furthermore, considering multiple quantum clocks, one finds that the reaction of the spacetime to their presence acts to limit the joint measurability of time along nearby worldlines~\cite{ruiz2017entanglement}.

In this work we consider a semiclassical version of the light-clock, whose mirrors follow classical trajectories but whose light rays are quantized. We identify the time as measured by this clock as that experienced by an accompanying observer, a quantity which deviates from the observer's proper time. We generalize the model introduced in~\cite{lindkvist2014twin}, including the effect of spacetime curvature (a challenge described in~\cite{lock2017relativistic}), and apply this to the example of a clock which falls a certain distance in a gravitational field. The model is detailed in section~\ref{sModel}, showing the state-independence of the clock time for Gaussian states, and the separation of the clock time into classical and quantum contributions. Section~\ref{sResults} presents the falling-clock scenario, a comparison with a corresponding classical observer, and numerical investigations into quantum effects on the clock time, followed by a discussion of our results in section~\ref{sDisc}.


\section{A quantum model of the light-clock} \label{sModel}

Neglecting polarisation effects, we can model the electromagnetic field by a massless real scalar field $\Phi$~\cite{friis2013scalar}, and we work in $1+1$ dimensions. Assuming the mirrors to be perfectly reflecting at all frequencies, they are modelled by the boundary conditions that the field vanishes at their location. The equation of motion for $\Phi$ is $\square\Phi=0$ where the d'Alembertian is given by ${\square:=(\sqrt{-g})^{-1}\partial_{\mu}\sqrt{-g}g^{\mu\nu}\partial_{\nu}}$ and $g:=\text{det}(g_{\mu\nu})$. For a stationary clock in a static spacetime, the field operator can be expanded as a sum of modes~\cite{birrell1984quantum}:
\begin{equation} \label{eTotalField}
\Phi = \sum_{m=1}^{\infty} \left[ a_{m} \phi_{m} +  a_{m}^\dag \phi_{m}^{*}  \right],
\end{equation}
where the $a_m$, satisfy the canonical commutation relations $[a_{m},a^{\dag}_{n}]=\delta_{mn}$, and the $\phi_{m}$ and the $\phi_{m}^{*} $ are positive and negative-frequency solutions to the equation of motion respectively. One can then define a vacuum state and a Fock space in the usual way.

In a classical light-clock, a ``tick'' corresponds to one round-trip of a light beam. Here, as in~\cite{lindkvist2014twin,lindkvist2015motion} we use the oscillations of the phase of the first mode ($m=1$) of a given quantum state. In the same manner as the classical clock, this can be converted into a quantity with dimensions of time using the distance betweem the mirrors and the speed of light. For simplicity, and to avoid the complications associated with defining an operator corresponding to phase (e.g. its non-uniqueness, corresponding to different possible measurement schemes - see section~III.5 in~\cite{busch1997operational}), we restrict ourselves to Gaussian quantum states and use an optical phase space description. Examples of Gaussian states include coherent states (sometimes referred to as the most classical quantum states) and squeezed states. Defining the quadrature operators of mode $m=1$ by $q:=\frac{1}{\sqrt{2}}\left( a_{1}+a_{1}^{\dag}\right)$ and $p:=-\frac{i}{\sqrt{2}}\left( a_{1}-a_{1}^{\dag}\right)$, the mean phase $\theta$ of a state is then given by
\begin{equation} \label{ePhaseBasic}
\tan \theta = \frac{\langle p \rangle}{\langle q \rangle},
\end{equation}
where the angular brackets denote the expectation value of an operator with respect to the state. Since phase is a relative quantity, there is a freedom in how the quadrature operators are defined; multiplying $a_{1}$ by an arbitrary reference phase in the definitions above acts to shift the phase $\theta$ in~(\ref{ePhaseBasic}). Setting $\theta=0$, for example, is equivalent to choosing this reference phase such that $\langle p \rangle=0$.

We will use Bogoliubov transformations to describe the evolution of a clock. These are linear transformations from one basis of mode solutions to another, or equivalently, from one set of creation and annihilation operators to another:
\begin{equation} \label{eBogo}
b_{m}=\sum_{n} \left( \alpha^{*}_{mn}a_{n}-\beta^{*}_{mn}a^{\dag}_{n} \right) ,
\end{equation}
where the $b_{m}$ are the new annihilation operators, and the new creation operators are evidently obtained by taking the Hermitian conjugate of~(\ref{eBogo}). At the level of the quantum state, this transformation is represented by a (Gaussian) unitary operator, i.e. a transformation preserving the Gaussian character of the state~\cite{adesso2014continuous}. The $\alpha_{mn}$ and $\beta_{mn}$ are known as Bogoliubov coefficients, and respectively quantify the effects of mode mixing and particle creation. Given a Gaussian initial state with first moments  $\langle x \rangle_{0}$ and $\langle p \rangle_{0}$, the phase of a clock after a transformation is given by~\cite{lindkvist2015motion}
\begin{equation} \label{ePhaseTrans}
\tan \theta = \frac{-\text{Im} \left( \alpha_{11} - \beta_{11} \right) \langle x \rangle_{0} + \text{Re} \left( \alpha_{11} + \beta_{11} \right) \langle p \rangle_{0}}{ \text{Re} \left( \alpha_{11} - \beta_{11} \right) \langle x \rangle_{0} + \text{Im} \left( \alpha_{11} + \beta_{11} \right) \langle p \rangle_{0}} .
\end{equation}
From~(\ref{ePhaseTrans}), we see that if we define the quadrature operators such that the clock is initialized with zero phase (i.e. $\langle p \rangle_{0}=0$), then the final phase no longer depends on the initial state, but only on the Bogoliubov coefficients, which in turn depend on the mode solutions to the field equation and the motion of the boundaries. The time as measured by the clock is then uniquely determined by the background spacetime metric and the clock motion. As a consequence, the relativistic phase shifts predicted in~\cite{lindkvist2014twin} hold for all Gaussian states, not only the coherent state used therein.

We consider two light-clocks, with their phases both initialized at zero. The reference clock, labelled $\text{A}$, remains at rest (with respect to the stationary spacetime), and the other clock, labelled $\text{B}$, undergoes a finite period of motion. To each clock we associate a classical observer, following one of the clock-mirrors along its trajectory. We will examine how the readings of the two clocks differ, and contrast this with the difference in the proper time of the corresponding classical observers. We use coordinates in which the metric is conformally flat, denoted by $(t,x)$, as bookkeeping coordinates (i.e. we do not associate them to an observer), and assume the motion of clock $\text{B}$ to take place in the interval $0<t<T$. The trajectories of the two mirrors of clock $\text{B}$ are denoted $x_{1}(t)$ and $x_{2}(t)$ (where $x_{1}(t)<x_{2}(t)$), and the instantaneous frequency (with respect to $t$) of mode~$m$ of this clock is therefore $\omega_{m}(t)=c \, m \pi / \left[ x_{2}(t)-x_{1}(t) \right]$. Assuming the two clocks to initially have the same length (in the $x$-coordinate), the mode frequencies of the stationary clock, $\text{A}$, are then $\omega_{m}^\text{A}:=\omega_{m}(0)$, and the free evolution of its field state corresponds to the Bogoliubov coefficients $\alpha_{mn}^{\text{A}}=\exp\left[i \omega_{m}^\text{A} T \right] \delta_{mn}$ and $\beta_{mn}^\text{A}=0$, giving a phase of $\theta_\text{A}=-\omega_{1}^\text{A}T$. The minus sign is a consequence of the standard definition of positive-frequency modes~\cite{birrell1984quantum}.
 
We then need to determine the Bogoliubov coefficients for clock $\text{B}$'s  state transformation. The separation of $\Phi$ into well-defined positive and negative-frequency modes in~(\ref{eTotalField}) is not possible for general boundary motion~\cite{fulling1976radiation}. We overcome this using the method introduced in~\cite{lock2017dynamical}, which allows us to calculate the Bogoliubov coefficients mapping the field solutions before the motion to those after it, and therefore determine the evolution of the quantum state due to the motion. Working to second order in the velocities $d x_{j}/dt$, one obtains Bogoliubov coefficients for clock $\text{B}$ which are of the form
\begin{subequations} \label{eBogoExpans}
\begin{align}
\alpha_{mn}^\text{B}=& e^{-i m \, \theta_{\text{B}}^{\text{Cl}}} \left[ \delta_{mn} + \bar{\alpha}_{mn}^{(1)}+\bar{\alpha}_{mn}^{(2)} \right]  \\
\beta_{mn}^\text{B}=& e^{-i m \, \theta_{\text{B}}^{\text{Cl}}} \left[ \bar{\beta}_{mn}^{(1)}+\bar{\beta}_{mn}^{(2)} \right] ,
\end{align}
\end{subequations}
where $\theta_{\text{B}}^{\text{Cl}}:=-\int_{0}^{T} \mathrm{d} t \, \omega_{1}(t)$ is the phase accrued by a classical oscillator with a time-dependent frequency, and where a term's superscript in parentheses denotes its order in the mirror velocities. The full expression of each term in~(\ref{eBogoExpans}) is given in appendix~\ref{a1}. The phase of clock~$\text{B}$ is then given by
\begin{equation} \label{ePhaseClQu}
\theta_{\text{B}} = \theta_{\text{B}}^{\text{Cl}}+ \theta_{\text{B}}^\text{Qu} 
\end{equation}
with
\begin{equation} \label{ePhaseDiffQu}
\begin{aligned}
\theta_{\text{B}}^\text{Qu}  :=& - \left[ \text{Im} \left( \bar{\alpha}_{11}^{(1)} - \bar{\beta}_{11}^{(1)} \right) + \text{Im} \left( \bar{\alpha}_{11}^{(2)} - \bar{\beta}_{11}^{(2)} \right) \right] \\
& + \, \text{Re} \left( \bar{\alpha}_{11}^{(1)} - \bar{\beta}_{11}^{(1)} \right) \text{Im} \left( \bar{\alpha}_{11}^{(1)} - \bar{\beta}_{11}^{(1)} \right) .
\end{aligned} 
\end{equation}
In~(\ref{ePhaseClQu}) we see that $\theta_{\text{B}}$ consists of two contributions: the phase of a classical variable-frequency oscillator, plus an effect arising purely from the motion-induced transformation of the clock's quantum state. In the next section, we explore these two effects with a specific example.

We have described how to find the phase accrued by both clocks between bookkeeping time $t=0$ and $t=T$. In~\cite{lindkvist2014twin}, one could convert phases to quantities with dimensions of time by dividing by the frequency of the clock mode (as is done in the measurement of time by atomic clocks~\cite{poli2013optical}). This was possible because the frequency is the same (to within the approximation used) for both clocks. In curved spacetime however, matters are more complicated. Generally, the frequency of clock~$\text{B}$'s reference mode will vary with respect to the bookkeeping time $t$, and therefore with respect to clock $\text{A}$, throughout the motion. For an observer carrying clock~$\text{B}$ to determine how this frequency varies, they would need to know the trajectories of the clock's boundaries, as well as the value of $T$, which is not in keeping with the principle of operationalism unless we specify the manner by which the observer obtains knowledge of this time.

To obtain an operational comparison of the two clocks, we compare their phases directly, considering the fractional phase difference
\begin{equation}
\mathcal{F}_\theta := \frac{\theta_{\text{B}} - \theta_\text{A}}{\theta_\text{A}}.
\end{equation}
Given the separation described in~(\ref{ePhaseClQu}), $\mathcal{F}_\theta$ likewise separates into two terms $\mathcal{F}_\theta = \mathcal{F}_{\theta}^{\text{Cl}} + \mathcal{F}_{\theta}^\text{Qu}$, with $\mathcal{F}_{\theta}^{\text{Cl}} := \left( \theta_{\text{B}}^{\text{Cl}} - \theta_\text{A} \right) / \theta_\text{A}$ and $\mathcal{F}_{\theta}^\text{Qu} := \theta_{\text{B}}^\text{Qu} / \theta_\text{A}$. Denoting the proper times of the pointlike, ideal observers associated with clocks $\text{A}$ and $\text{B}$ by $\tau_\text{A}$ and $\tau_{\text{B}}$ respectively, we consider the fractional difference between these two
\begin{equation}
\mathcal{F}_\tau := \frac{\tau_{\text{B}} - \tau_\text{A}}{\tau_\text{A}},
\end{equation}
which can be compared with the fractional phase difference between the two light-clocks, as we do in the following example.


\section{A falling quantum clock}  \label{sResults}

\subsection{The scenario}\label{sScenario}

\begin{figure}[h]
  \centering
  \includegraphics[width=0.65\textwidth]{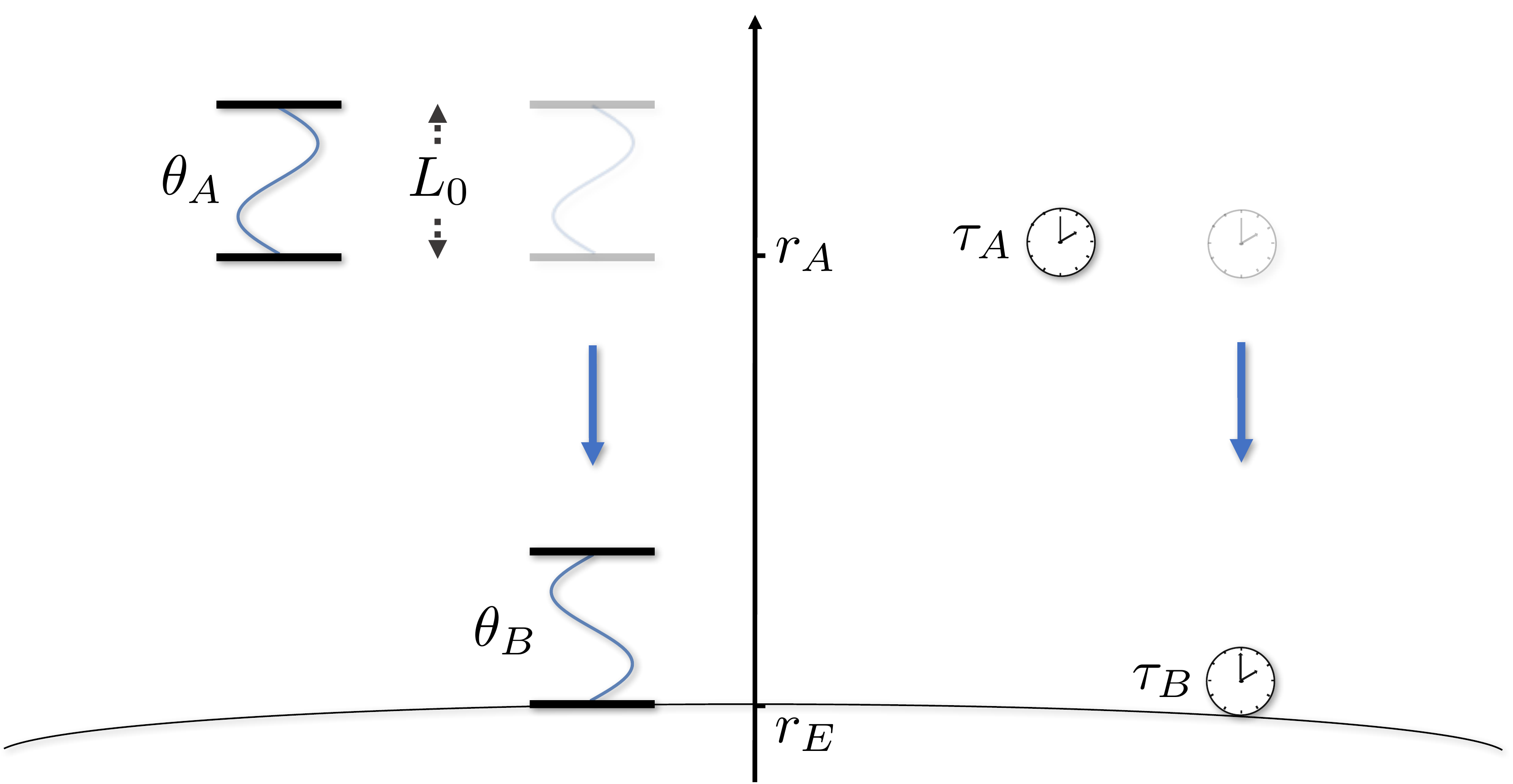}
  \caption{An illustration of the scenario, showing the initial and final heights of the light-clocks (left) and the associated ideal, pointlike clocks (right). We take $r_\text{E}=6367$ km and $r_\text{A}=r_\text{E}+100$~m.} \label{fCartoon}
\end{figure}

We now consider an example using the model discussed in section~\ref{sModel}. In particular, we wish to examine the magnitude of the difference between the two clock times in a parameter regime accessible on Earth. We consider the scenario where clock $\text{A}$ is kept fixed 110m above the Earth's surface and clock $\text{B}$ falls freely from that height to the surface, comparing the fall-time as measured by the two clocks. This is illustrated in figure~\ref{fCartoon}. We consider both of clock~$\text{B}$'s mirrors to fall freely, rather than being fixed by some rigid support. The choice of height, $110$~m, is based on the size of the Bremen Drop Tower at the Center of Applied Space Technology and Microgravity (ZARM)~\cite{ZARM}. 

We model the spacetime around the Earth using the Schwarzschild metric
\begin{equation} \label{eSchwarzInfos}
ds^{2} = -f(r) dt^{2} + \frac{1}{f(r)} dr^{2},
\end{equation}
with $f(r) :=1-r_\text{S}/r$ and $r_\text{S}:=2 G M$, where $M$ is the mass of the Earth (except where stated otherwise). For each clock, we take their corresponding observer to follow the lower mirror (the one with smaller $r$). Observer $\text{A}$, at position $r_\text{A}$, experiences the proper time $\tau_\text{A}=\sqrt{f(r_\text{A})}T$ during the motion. The trajectory of an observer falling freely from rest at $(t,r)=(0,r_{0})$ satisfies
\begin{equation} \label{eSchwarzInGeodesic}
\frac{d}{d \tau_\text{d}} (t,r) = \left(  \frac{\sqrt{f(r_{0})}}{f(r)} ,  - \sqrt{f(r_{0})-f(r)} \right),
\end{equation}
with the proper time $\tau_\text{d}$ of this ``drip'' observer defined such that $\tau_\text{d}=0$ at $t=0$. Solving these equations for two different values of $r_{0}$ gives the trajectories followed by the mirrors of clock~$\text{B}$: specifically $r_{0}=r_\text{A}$ for the bottom mirror and $r_{0}=r_\text{A}+L_{0}$ for the top one, where $L_{0}$ is the initial clock size. Introducing the tortoise coordinate ${x:=r+r_\text{S} \ln \left\vert r / r_\text{S}-1 \right\vert}$, we have the conformally flat coordinate system $(t,x)$ necessary for computing the transformation of clock~$\text{B}$'s phase. The proper time $\tau_{\text{B}}$ experienced by observer $\text{B}$ is given by solving~(\ref{eSchwarzInGeodesic}) for $\tau_\text{d}$ with $r_{0}=r_\text{A}$.

\subsection{Classical light-clocks and proper times} \label{sClassical}

We now examine the classical part of the phase discrepancy between clocks $\text{A}$ and $\text{B}$, comparing this with the proper times of the corresponding pointlike observers. First, we can gain some insight by solving for the mirror trajectories to second order in $r_\text{S}/r_\text{A}$ and $L_{0}/r_\text{A}$ to obtain the classical phase discrepancy between the two clocks
\begin{equation} \label{eClassicalFracChange}
\mathcal{F}_{\theta}^{\text{Cl}} = \frac{1}{2} \left( \frac{a_\text{A} L_{0}}{c^{2}} - \frac{1}{3} \frac{r_\text{S}}{r_\text{A}}  \right) \left( \frac{c T}{r_\text{A}} \right)^{2} ,
\end{equation}
where $a_\text{A}:=r_\text{S}/2 \sqrt{f(r_\text{A})} r_\text{A}^{2}$ is the proper acceleration felt by an observer at rest at $r_\text{A}$ (the starting point of $\text{B}$'s trajectory). The component proportional to  $a_\text{A} L_{0}/2c^2$ is reminiscent of the case of an accelerating light clock in flat spacetime (see Exercise 3.10 in~\cite{rindler2006relativity}), though here it is the stationary clock which undergoes proper acceleration. In the scenario we consider here, $a_\text{A}/c^{2} \sim 10^{-16}\text{m}^{-1}$, while $r_\text{S}/r_\text{A}\sim10^{-9}$, and therefore~(\ref{eClassicalFracChange}) tells us that the classical discrepancy is extremely insensitive to changes in length in this regime. If the classical light-clock were ideal, we would have $\mathcal{F}_{\theta}^{\text{Cl}} = \mathcal{F}_{\tau}$, which is evidently not the case. Rather, we have
\begin{equation}
\mathcal{F}_{\theta}^{\text{Cl}} =\left( 1 - \frac{a_{\text{tide}} T^{2}}{6 L_{0}} \right) \mathcal{F}_{\tau}
\end{equation}
asymptotically as $L_{0} \to 0$, where $a_{\text{tide}}$ is the classical tidal acceleration between $r_\text{A}$ and $r_\text{A}+L_{0}$. This dependence on tidal acceleration contrasts with the flat-spacetime case considered in~\cite{lindkvist2014twin}.

\begin{figure}[h]
  \centering
  \includegraphics[width=0.65\textwidth]{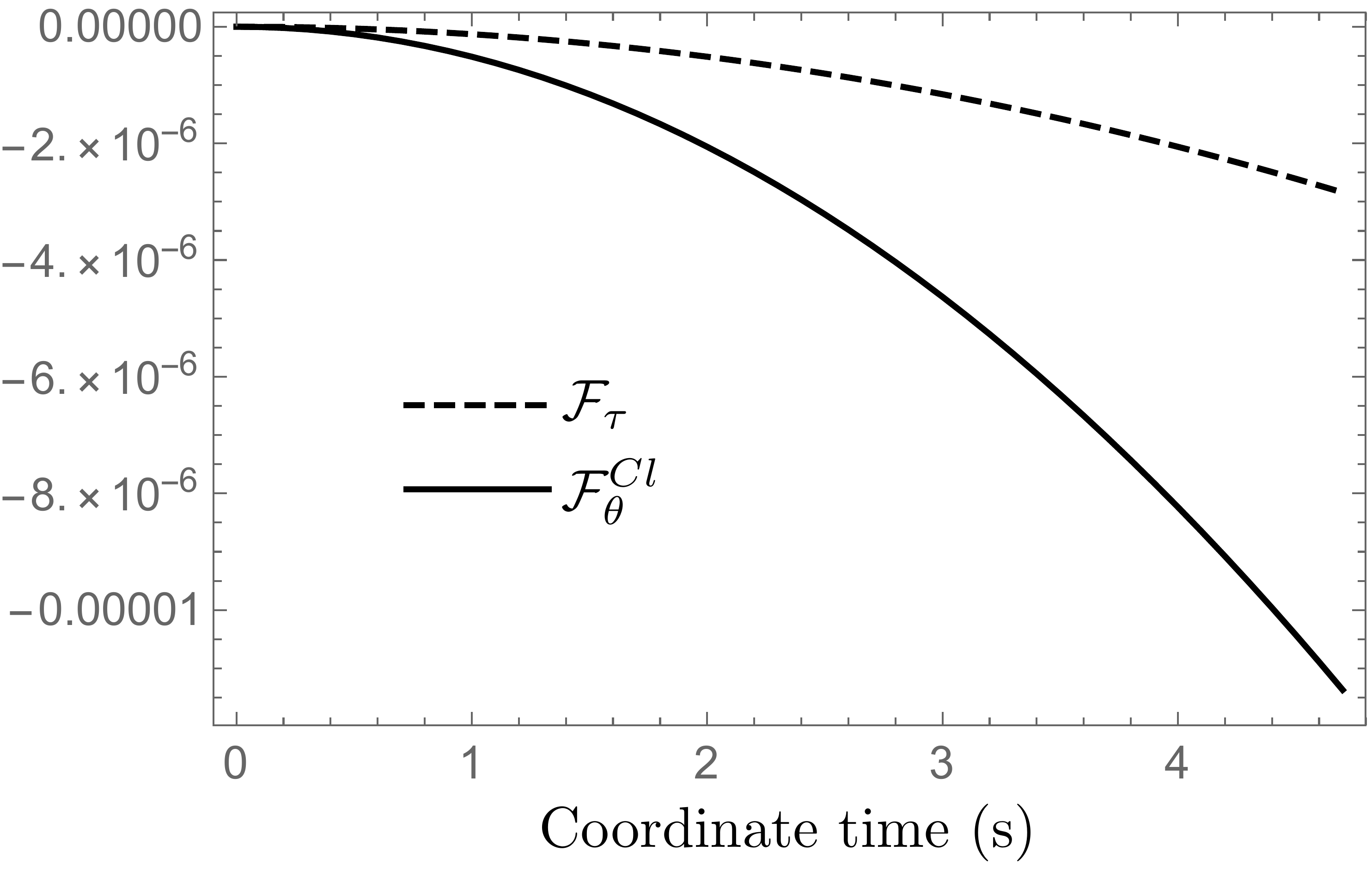}
  \caption{The fractional difference between the readings of dropped and stationary clocks in the case of pointlike, ideal clocks (dashed line), and classical light-clocks (solid line), as a function of $t$. Their negative values indicate that the dropped clocks experience less time passing than the stationary ones. Curves shown for $L_{0}=1$~m and a fall of $110$~m to the surface of the Earth.} \label{fClockPointComp}
\end{figure}

The evolution of $\mathcal{F}_{\theta}^{\text{Cl}}$ and $\mathcal{F}_{\tau}$ during the fall are given in figure~\ref{fClockPointComp}. We can see that the classical light-clock exhibits a greater fractional discrepancy compared to the ideal case, and that this effect is of considerable magnitude. While this is given in terms of the bookkeeping coordinate $t$, we recall that this coordinate is simply the proper time of observer $\text{A}$ scaled by the factor~$\sqrt{f(r_\text{A})}$.

\subsection{Quantum effects on the falling light-clock} \label{sQuantum}

\begin{figure}[h]
  \centering
\begin{subfigure}[b]{0.65\textwidth}
  \includegraphics[width=0.98\textwidth]{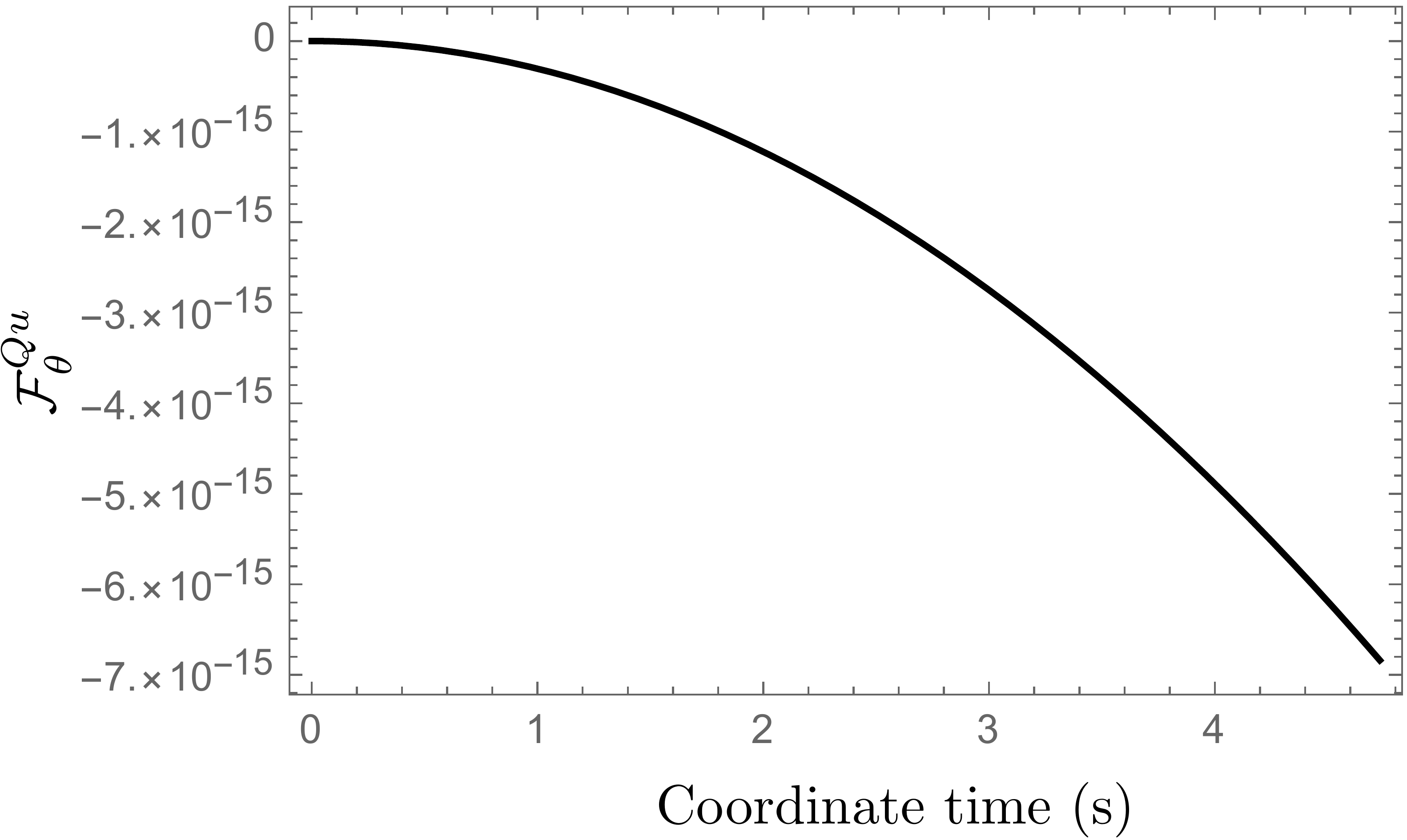}
        \caption{The fractional quantum effect for $L_{0}=1$~m, though the behaviour on this scale is effectively length-independent (see main text).} \label{fQuFracTime}
    \end{subfigure} \\ \vspace{6mm}
    ~
    \begin{subfigure}[b]{0.65\textwidth}
  \includegraphics[width=0.98\textwidth]{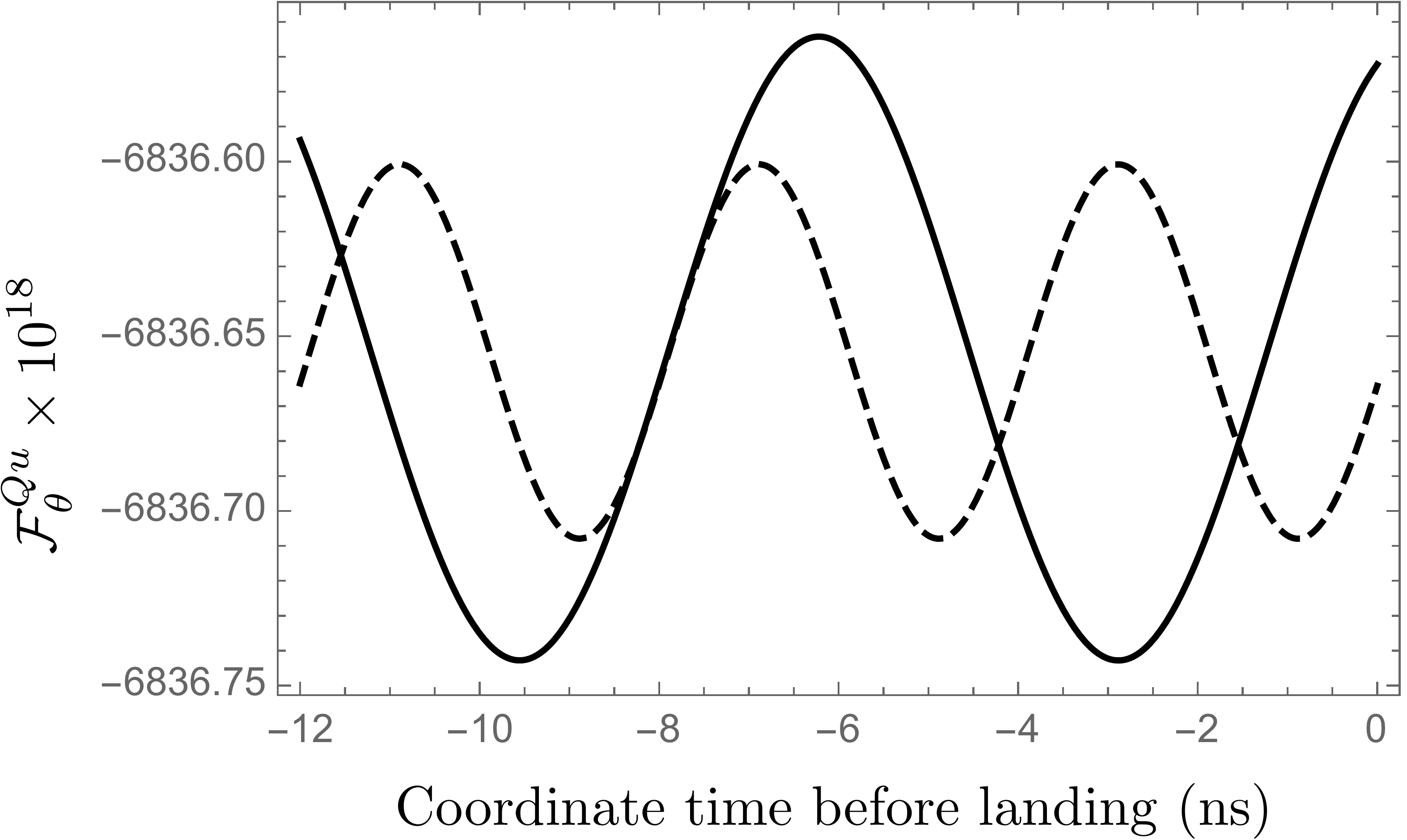}
        \caption{Small-scale oscillations (magnified by $10^{18}$) towards the end of clock $\text{B}$'s fall, shown for $L_{0}=1$~m (solid line) and $L_{0}=0.6$~m (dashed line).} \label{fOscil}
    \end{subfigure} 
  \caption{The (a) overall and (b) small-time-scale quantum contribution ($\mathcal{F}_{\theta}^\text{Qu}:=\theta_{\text{B}}^\text{Qu}/\theta_\text{A}$) to the fractional phase discrepancy between the clocks during a fall of $110$~m to the surface of the Earth. This contribution acts to increase the difference in phase between the two clocks. For comparison, $\mathcal{F}_{\theta}^{\text{Cl}}\approx-10^{-5}$ at the end of the motion.}
\end{figure}

The quantum contribution to the fractional phase discrepancy between the light-clocks was given in Section~\ref{sModel} by~$\mathcal{F}_{\theta}^\text{Qu}$. The behaviour of this quantity during the fall is shown in figure~\ref{fQuFracTime}. There we see that the motion-induced change in the quantum state acts to increase the magnitude of $\mathcal{F}_{\theta}$ (recall $\mathcal{F}_{\theta}^{\text{Cl}} \leq 0$). We do not find a closed-form expression for $\mathcal{F}_{\theta}^\text{Qu}$ like the one in~(\ref{eClassicalFracChange}), but numerically we find the same insensitivity to clock size. Specifically, considering initial lengths from $0.1$~$\upmu$m to $10$~m (in the $r$-coordinate), we find that the curve in figure~\ref{fQuFracTime} does not noticeably change. Examining the quantum effect on a much smaller scale reveals an oscillatory behaviour (figure~\ref{fOscil}). These oscillations occur at a frequency comparable to that of the clock mode, and are then affected by changes in clock size, in contrast with the large-scale behaviour in figure~\ref{fQuFracTime}. Specifically, they increase in frequency but also decrease in amplitude as we consider smaller clocks. figure~\ref{fOscil} illustrates this behaviour for two different values of initial length, finding oscillations of amplitude $\sim 10^{-19}$ for initial lengths on the order of $1$~m.

\subsection{Curvature dependence}

\begin{figure}[H]
  \centering
  \includegraphics[width=0.65\textwidth]{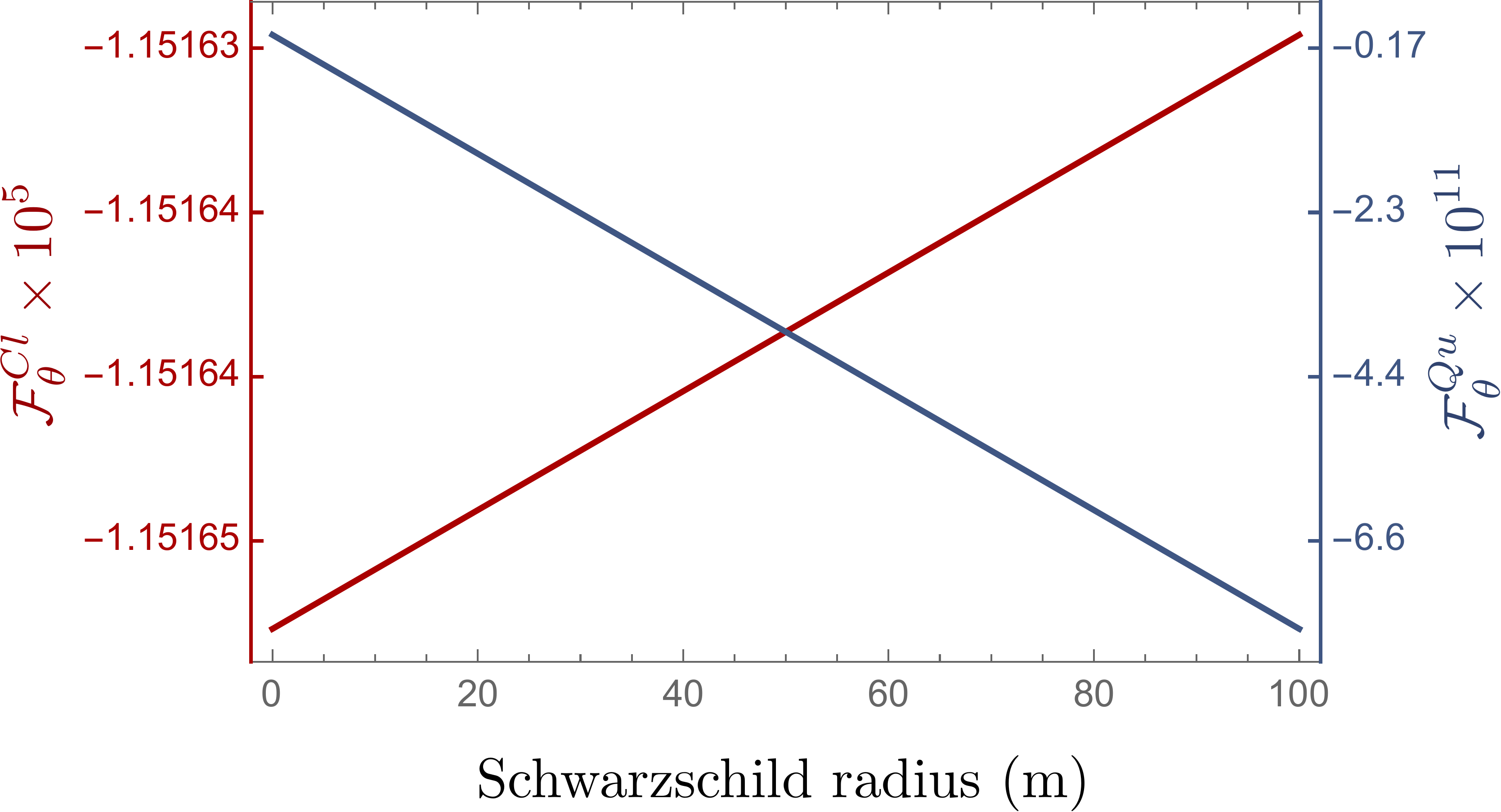}
  \caption{The approximately linear behaviour of the fractional classical ($\mathcal{F}_{\theta}^{\text{Cl}}$) and quantum ($\mathcal{F}_{\theta}^\text{Qu}$) contributions to the fractional phase difference $\mathcal{F}_{\theta}$ between $\text{A}$ and $\text{B}$, as a function of the spacetime curvature (i.e. Schwarzschild radius), after a fall of $110$~m to the surface of the Earth. Note that, since the phases are negative, the rising (falling) curve is decreasing (increasing) in magnitude. For comparison, the Schwarzschild radius of the Earth is $r_\text{S}\sim1$~cm.} \label{fRSDep}
\end{figure}

Though we have been primarily interested in the parameter regime corresponding to the Earth, it is interesting to consider how our results change with the curvature (as quantified by $r_\text{S}$). We consider the range $0<r_\text{S}<100$~m, while continuing to use the same $r_\text{A}$ as before (i.e. fixed at $110$~m higher than the radius of the Earth). The results are shown in figure~\ref{fRSDep}. The overall fall time decreases with increasing curvature, an effect which wins out over tidal forces to decrease the magnitude of $\mathcal{F}_{\theta}^{\text{Cl}}$. This approximate linearity of this decrease in the regime considered here is shown by the red curve in figure~\ref{fRSDep}. This behaviour is not surprising, given~(\ref{eClassicalFracChange}). On the other hand, we find a seemingly linear increase in the magnitude of the fractional quantum effect $\mathcal{F}_{\theta}^\text{Qu}$. (the blue curve in figure~\ref{fRSDep}).


\section{Discussion} \label{sDisc}

While the quantum effect on the clock phase is small in a terrestrial drop-tower experiment, being on the order of $10^{-15}$ times the stationary clock's phase (figure~\ref{fQuFracTime}), it is not insignificant, particularly given the fractional frequency instabilities of $10^{-18}$ achievable by modern optical clocks~\cite{hinkley2013atomic}. An attempt to observe it would however face serious difficulties, including the maintenance of stability as the clock undergoes a change in acceleration, and the short integration time afforded by the fall. Furthermore, one would presumably require the two mirrors be fixed to a support, whose rigidity must be taken into account~\cite{ratzel2018frequency}.

The oscillating aspect of the quantum effect distinguishes it qualitatively from the classical one, raising interesting questions from an operationalist perspective. A local time reference constructed via the moving clock would exhibit a many-to-one relationship with respect to the stationary one, and allow for the possibility that the order of timelike separated events to be different for observers $\text{A}$ and $\text{B}$, in contrast with the usual causal structure of relativity. That quantum mechanics may give rise to new causal structures is discussed in e.g.~\cite{oreshkov2012quantum}. One could instead interpret the oscillating aspect as the breakdown of clock~$\text{B}$'s utility for constructing a time reference, and hence the emergence of a minimum meaningful time-scale in this model due to the motion. This is distinct from the usual notion of quantum uncertainty resulting from the variation in the measurements performed on quantum states. The decrease in amplitude of the oscillations with decreasing clock size could then be interpreted as an increase in the resolution of the clock as the frequency (i.e. energy) of the clock mode increases.

One could, as in the classical case, argue that clock~$\text{B}$'s deviation from ideality is an artifact, and can be subtracted by measuring the trajectory of its mirrors. This is not without its conceptual issues, as quantum theory demands that there be some uncertainty associated with this measurement, hindering an agent's efforts to perform this ``recalibration''. This would be particularly problematic if the uncertainty associated with the recalibration is of non-negligible magnitude compared to the quantum effect on the clock time. We cannot analyse this within the current formalism however, which treats the clock mirrors as classical objects. One might then argue that the effects we predict here show that we are simply using the ``wrong'' system as a clock. Such an argument must be followed by specifying the ``right'' system to use as a clock, since even atomic clocks are subject to deviations from ideality when subjected to tidal forces (see e.g.~\cite{misner1973gravitation}, p.396). 

There are a number of other limitations to our approach. We have considered only Gaussian field states in $1+1$ dimensions, neglecting polarisation. We have not considered the process by which the clock phase is measured. There is necessarily some uncertainty associated with this, determined by the clock's quantum state~\cite{lindkvist2015motion}. We have likewise not considered the problem of comparing the phases of two spatially separated systems in a curved background, which may be accompanied by its own operational issues, such as perhaps requiring observers to have accurate knowledge of the spacetime metric, and of each system's position. Moreover, the phase is periodic, and therefore a system for counting the oscillations must be included. The requirement that this multipartite system function continuously leads to a number of quantum and thermodynamic considerations (see e.g.~\cite{erker2017autonomous,erker2014quantum}). While consideration of any of these aspects may numerically affect the results above, it seems very unlikely that any of them will be able to restore ideality to the clock.


\section{Conclusion}

We investigated a model of a quantum light-clock which moves in a stationary spacetime. It was shown for Gaussian field states that the mean phase shift (and therefore readout) of such a clock depends, only the transformation applied to that state, which itself depends only on the spacetime and the motion, as is the case for an ideal classical clock. Assuming low-velocities (in relevant coordinates), we found that the phase of a clock after a period of motion separates into the sum of a classical part, incorporating the changing frequency (i.e. photon round-trip time) of the clock, and a quantum part, arising from the transformation of the quantum state due to the motion. We then presented a numerical investigation of a scenario where one clock is held at a constant height above the surface of the Earth (whose gravitational field we model using the Schwarzschild metric) and the other is dropped to the surface. We first examined the classical part of the phase discrepancy between the clocks. Comparing this discrepancy with the time dilation experienced between two corresponding ideal classical observers, we found that the former is greater in magnitude. The quantum effect further delayed the dropped clock. Considering this effect on a smaller time-scale, one finds a slight oscillatory behaviour, varying with clock-size. Finally, moving away from a terrestrial setting, we found the classical and quantum effects respectively decrease and increase in magnitude with increasing curvature, raising the possibility of a regime in which quantum effects become significant.


\section*{Acknowledgments}
The authors would like to thank Luis C.~Barbado and Tupac Bravo for useful discussions. MPEL acknowledges support from the EPSRC via the Controlled Quantum Dynamics CDT (EP/G037043/1), as well as the ESQ Discovery Grant (ESQ-Projekts0003X2) of the Austrian Academy of Sciences (\"{O}AW), and IF acknowledges support from FQXi via the `Physics of the observer' award `Quantum Observers in a Relativistic World', as well as The Penrose Institute.


\appendix

\section{Clock B's Bogoliubov coefficients}\label{a1}
Writing the Bogoliubov coefficients corresponding to clock~$\text{B}$'s transformation as $\alpha_{mn}^{B}= \alpha_{mn}^{(0)}+ \alpha_{mn}^{(1)}+\alpha_{mn}^{(2)}$ and ${\beta_{mn}^{B}=\beta_{mn}^{(1)}+\beta_{mn}^{(2)}}$, the individual terms are given by~\cite{lock2017dynamical}
\begin{subequations} \label{eBogosFull}
\begin{align}
\alpha_{mn}^{(0)} &= e^{i \Theta_{m}(T)} \delta_{mn}	\\
\alpha_{mn}^{(1)} &= e^{i \Theta_{m}(T)} \sum_{j=1}^{2} \int_{0}^{T} \mathrm{d} t \, A^{j}_{mn}  e^{-i \left[ \Theta_{m}(t) - \Theta_{n}(t) \right]} \frac{d x_{j}}{d t} 	\\
\alpha_{mn}^{(2)} &= e^{i \Theta_{m}(T)} \sum_{j,k=1}^{2} \int_{0}^{T} \mathrm{d} t_{2} \int_{0}^{t_2} \mathrm{d} t_{1} \, C_{mn}^{jk}(t_{1},t_{2})  e^{-i[\Theta_m (t_2) - \Theta_n (t_1)]} \frac{d x_{j}}{d t_2} \frac{d x_{k}}{d t_1} \\
\beta_{mn}^{(1)} &=  e^{i \Theta_{m}(T)} \sum_{j=1}^{2} \int_{0}^{T} \mathrm{d} t \, B^{j}_{mn}  e^{-i \left[ \Theta_{m}(t) + \Theta_{n}(t) \right]} \frac{d x_{j}}{d t}	\\
\beta_{mn}^{(2)} &=  e^{i \Theta_{m}(T)} \sum_{j,k=1}^{2} \int_{0}^{T} \mathrm{d} t_{2} \int_{0}^{t_2} \mathrm{d} t_{1} \, D_{mn}^{jk}(t_{1},t_{2}) e^{-i[\Theta_m (t_2) + \Theta_n (t_1)]} \frac{d x_{j}}{d t_2} \frac{d x_{k}}{d t_1} ,
\end{align}
\end{subequations}
where $j$ and $k$ label the two mirrors of the clock, and with the following definitions
\begin{subequations}
\begin{align}
\Theta_{m}(t)&:=\int_{0}^{t} \mathrm{d} t' \omega_{m}(t') 	\\
A^{j}_{mn} & := \left( \frac{\partial \phi_{m}}{\partial x_{j}} , \phi_{n} \right)  \\
B^{j}_{mn} & :=-\left( \frac{\partial \phi_{m}}{\partial x_{j}} , \phi_{n}^{*} \right) \\
C_{mn}^{jk}(t_{1},t_{2}) &:=  \sum_{p} \left[ A^{j}_{mp}A^{k}_{pn} e^{i [\Theta_{p}(t_2) -\Theta_{p}(t_1)]} + B^{j}_{mp}B^{k}_{pn} e^{-i [\Theta_{p}(t_2) -\Theta_{p}(t_1)]} \right] \\
D_{mn}^{jk}(t_{1},t_{2}) &:= \sum_{p} \left[ A^{j}_{mp}B^{k}_{pn} e^{i [\Theta_{p}(t_2) -\Theta_{p}(t_1)]} + B^{j}_{mp}A^{k}_{pn} e^{-i [\Theta_{p}(t_2) -\Theta_{p}(t_1)]} \right] ,
\end{align}
\end{subequations}
where $\left( \cdot \, , \cdot\right)$ denotes the usual inner product between solutions to the Klein-Gordon equation~\cite{birrell1984quantum}. The forms of $\bar{\alpha}_{mn}^{(1)}$, $\bar{\alpha}_{mn}^{(2)}$, $\bar{\beta}_{mn}^{(1)}$ and $\bar{\beta}_{mn}^{(2)}$ in~(\ref{eBogoExpans}) can then be obtained by comparison with~(\ref{eBogosFull}).


\normalem
\bibliographystyle{naturemag}
\bibliography{fallingclock_bibtex}

\end{document}